# Petascale Computational Systems:
# Balanced CyberInfrastructure in a Data-Centric World

Gordon Bell[1], Jim Gray[1] and Alex Szalay[2]
1. Microsoft Research
2. The Johns Hopkins University
GBell@Microsoft.com, Gray@Microsoft.com, Szalay@jhu.edu
September 2005

**Abstract:** Computational science is changing to be *data intensive*. Super-Computers must be *balanced systems*; not just CPU farms but also petascale IO and networking arrays. Anyone building CyberInfrastructure should allocate resources to support a *balanced Tier-1 through Tier-3* design.

## Computational Science and Data Exploration

Computational Science is a new branch of most disciplines. A thousand years ago, science was primarily *empirical*. Over the last 500 years each discipline has grown a *theoretical* component. Theoretical models often motivate experiments and generalize our understanding. Today most disciplines have both empirical and theoretical branches. In the last 50 years, most disciplines have grown a third, *computational* branch (e.g. empirical, theoretical and computational ecology, or physics, or linguistics). Computational Science has meant simulation. It grew out of our inability to find closed form solutions for complex mathematical models. Computers can simulate these complex models.

Over the last few years Computational Science has been evolving to include information management. Scientists are faced with mountains of data that stem from four trends: (1) the flood of data from new scientific instruments driven by Moore's Law – doubling their data output every year or so; (2) the flood of data from simulations; (3) the ability to economically store petabytes of data online; and (4) the Internet and computational Grid that makes all these archives accessible to anyone anywhere exacerbating the replication, creation, and recreation of more data.

Acquisition, organization, query, and visualization tasks scale almost linearly with data volumes. By using parallelism, these problems can be solved within fixed times (minutes or hours). In contrast, most statistical analysis and data mining algorithms are nonlinear. Many tasks involve computing statistics among sets of data points in some metric space. Pair-algorithms on $N$ points scale as $N^2$. If the data increases a thousand fold, the work and time can grow by a factor of a million. Many clustering algorithms scale even worse. These algorithms are infeasible for terabyte-scale datasets.

## Computational Problems are Becoming Data-Centric

Next generation computational systems and science instruments will generate petascale information stores. The computation systems will often be used to analyze these huge information stores. For example BaBar processes and reprocesses a petabyte of event data today. About 60% of the BaBar hardware budget is for storage and IO bandwidth [1]. The Atlas and CMS systems will have requirements at least 100x higher. The Large Scale Synoptic Telescope (LSST) has requirements in the same range: peta-ops of processing and tens of petabytes of storage.

SETI@home and similar projects show that one can do interesting science with an IO-poor environment – but those systems require that the CPU:IO ratio be 100,000 instructions per byte of IO or higher [2]. Cryptography, signal processing, and certain other problem domains have such cpu-intensive profiles, but most other scientific tasks are much more information intensive having CPU:IO ratios well below 10,000:1 in line with Amdahl's laws.

## Amdahl's Laws – Building Balanced Systems

System performance has been improving with Moore's law and it will continue as multi-core processors replace single processor chips and as memory hierarchies evolve. Within five years, we expect a simple, shared memory multiprocessor to deliver about ½ tera-ops. Much of the effort in building Beowulf clusters and the supercomputing centers has been focused on CPU-intensive TOP-500 rankings. Meanwhile, in most sciences the amount of data (both experimental and simulated) has been increasing even faster than Moore's law because the instruments are getting so much better and cheaper, and because storage costs have been improving much faster than Moore's law.

Gene Amdahl coined many rules of thumb for computer architects. Surprisingly, 40 years later, the rules still apply [3]:

**Amdahl's parallelism law:** *If a computation has a serial part S and a parallel component P, then the maximum speedup is S/(S+P).*
**Amdahl's balanced system law:** *A system needs a bit of IO per second per instruction per second: about 10 instructions per second implies a need for 1 byte of IO per second.*
**Amdahl's memory law:** $\alpha=1$: *that is the MB/MIPS ratio (called alpha ($\alpha$)), in a balanced system is 1.*
**Amdahl's IO law:** *Programs do one IO per 50,000 instructions*

In [3], it is shown that $\alpha$ has increased and that has caused a slight reduction in IO density, but these "laws" are still a decent rule-of-thumb. In addition to these Amdahl Laws, computer systems typically allocate comparable budget for RAM and for long-term storage (e.g. disk, tape.) The long-term storage is about one hundred times less expensive than RAM, so one gets considerably more storage. This 1:100 RAM:Disk capacity ratio and the Amdahl laws are captured in the following spreadsheet

| Table 1. Amdahl's Balanced System's Laws Applied to Various System Powers. | | | | | | |
|---|---|---|---|---|---|---|
| OPS | OPS | RAM | Disk IO Byte/s | Disks for that Bandwidth @ 100MB/s/disk | Disk Byte Capacity (100x RAM) | Disks for that Capacity @ 1TB/disk |
| giga | 1E+09 | gigabyte | 1E+08 | 1 | 1E+11 | 1 |
| tera | 1E+12 | terabyte | 1E+11 | 1,000 | 1E+14 | 100 |
| peta | 1E+15 | petabyte | 1E+14 | 1,000,000 | 1E+17 | 100,000 |
| exa | 1E+18 | exabyte | 1E+17 | 1,000,000,000 | 1E+20 | 100,000,000 |

Scaled to a peta-operations-per-second machine, these rules imply
- the parallel software to use that processor array,
- a petabyte of RAM ,
- 100 terabytes/sec of IO bandwidth and an IO fabric to support it,
- 1,000,000 disk devices to deliver that bandwidth (at 100 MB/s/disk),
- 100,000 disks storing 100 PB of data produced and consumed by this peta-ops machine (at 1TB/disk).

Indeed, these storage bandwidth numbers are daunting – a million disks to support the IO needs of a peta-scale processor. They indicate that the processors will be spending most of their time waiting for IO and memory – as is often the case today. There are precedents in such petabyte-scale distributed systems at Google, Yahoo!, and MSN Search [4]. Those systems have tens of thousands of processing nodes (approximately a peta-ops) and have ~ 100,000 locally attached disks to deliver the requisite bandwidth. These are not commodity systems, but they are in everyday use in many datacenters.

Once empirical or simulation data is captured, huge computational resources are needed to analyze the data and huge resources are needed to visualize the results of the data. Analysis tasks, involving petabytes of information require petascale storage and petascale IO bandwidth. Of course, the data needs to be reprocessed each time a new algorithm is developed and each time someone asks a fundamentally new question. That generates even more IO.

Even more importantly, to be useful, these databases require an ability to process the information at a semantic level – rather than just being a collection of bytes. The data needs to be curated with metadata, stored under a schema with a controlled vocabulary, and it needs to be indexed and organized for quick and efficient temporal, spatial, and associative search. These peta-scale database systems will be a major part of any successful petascale computational facility and will require substantial software investment.

## Data Locality—Bringing the Analysis to the Data

There is a well-defined cost associated with moving a byte of data across the Internet [4]. It is only worth moving the data to a remote computing facility if there are more than 100,000 CPU cycles per byte of data to perform the analysis. For less CPU intensive tasks it is better to co-locate the computation with the data. In a data-intensive world, where petabytes are common it is important to co-locate computing power with the databases rather than planning to move the data across the Internet to a "free" CPU. Moving data through the Internet is only "free" if the computation is more than $10^5$ instructions per byte – which is rare for a data analysis task. This poses a substantial software challenge. Much current middleware assumes that data movement is free. This is definitely not the case today, and it is unlikely in the future. Yes, we have been working on moving data from CERN to Pasadena at 1GB/s, but that data movement is expensive and so should be done only once if Pasadena wants to look at the data again.

## Computational Problem Sizes Follow a Power Law

The sizes of scientific computations depend on the product of many independent factors. Quantities formed as a product of independent random variables follow a lognormal distribution [5]. As a result, the cost of scientific computational problems has a power law (*1/f*) tail; the number of problems will be the same in each logarithmic interval. One can see this quite well in the current situation in US computing. Thirty years ago, supercomputers were the mainstay of computational science. Today the whole range is filled from Tier-1 supercomputers, to Tier-2 regional centers, to Tier-3 departmental Beowulf clusters, and to Tier-4, the single workstations. This 4-tier architecture reflects the problem size power-law.

## Building a Balanced CyberInfrastructure

What is the best allocation of cyberinfrastructure investments? There must certainly be two high-end Tier-1 centers that (1) allow competition, (2) allow design diversity, and (3) that leapfrog one another every two years. The Tier-1 facilities can only be built as a national priority. But, what should government agencies and industry do about the other tiers? They could make funding the Tier-2 and Tier-3 systems entirely the universities' responsibility—but that would be a mistake.

We believe that the available resources should be allocated to benefit the broadest cross-section of the scientific community. Given the power-law distribution of problem sizes [5], this means that part of funding agency resources should be spent on national, high-end Tier-1 centers at the petaop level; but that comparable amounts (about 50%) should be allocated to co-fund Tier-2 and Tier-3 centers. The resulting division would be a balanced allocation of resources with government agencies funding about ½ the Tier-2 and Tier-3 centers with institutions and other mission-agencies on a cost-sharing basis.

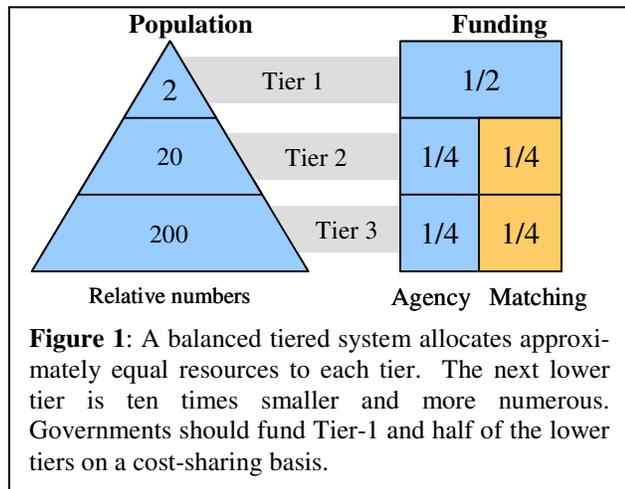

**Figure 1**: A balanced tiered system allocates approximately equal resources to each tier. The next lower tier is ten times smaller and more numerous. Governments should fund Tier-1 and half of the lower tiers on a cost-sharing basis.

It is interesting to note, that one of the most data intensive science projects to date, the CERN Large Hadron Collider, has adopted exactly such a multi-tiered architecture. The hierarchy of an increasing number of Tier-2 and Tier-3 analysis facilities provides impedance matching between the individual scientists and the huge Tier-1 data archives. At the same time, the Tier-2 and Tier-3 nodes provide complete replication of the Tier-1 datasets.

## An Example Tier-2 Node: Balanced Architecture and Cost Sharing

Most funding for Tier-2 and Tier-3 centers today split costs between the government and a university. In these arrangements, about ½ the costs fall to the University. It is difficult for Universities to get private donations towards computing resources because they depreciate so quickly. Donors generally prefer to donate money for buildings or endowed positions, which have a long term staying value. Therefore, government funding is crucial for Tier-2 and Tier-3 centers on a cost-sharing arrangement with the hosting institution.

To give a specific example, The Johns Hopkins University (JHU) is building a Tier-2 center. It received an NSF MRI grant to study turbulence through hydrodynamic simulations. It is interesting to see, how the principles outlined in this article worked in this case. JHU built a two-layer system; one is a traditional Beowulf cluster with 128 dual nodes connected with Myrinet, combined with a database layer of 100TB of storage spread over 12 database servers that store every time-step of a $1000^3$ resolution simulation. This arrangement provides a facility much closer to Amdahl's balance laws than a traditional Beowulf – a ~1.2 Teraflop has 100TB of disks as (300 disks) in the database layer, and 128 smaller disks in the compute layer. This is close to Amdahl's Law for disk capacity, and within a factor of 3 in bandwidth.

The NSF contributed about $480K towards this facility; JHU provided an initial cost sharing of $180K. The incremental power and cooling needs pushed the building above the original design threshold requiring an additional $300K to upgrade the hosting facilities. In addition, a 50% FTE systems person runs the facility. As a result, JHU's matching funds are 125%. Without the NSF contribution none of this would have happened. At the same time, due to the NSF award, the University allocated substantial additional resources, and the facility is now in use. We expect that other institutions have had similar experiences when setting up larger computing facilities: the price of computers is less than half the cost and the Universities can provide those infrastructure costs if NSF seeds the Tier-2 and Tier-3 centers.

## Summary

In summary we would like to emphasize the importance of building balanced systems, reflecting the needs of today's science, and also of building a balanced cyberinfrastructure. Placing all the financial resources at one end of the power-law (or lognormal) distribution would create an unnatural infrastructure, where the computing needs of most mid-scale scientific experiments will not be met. On the system level, placing all the focus on CPU harvesting will also tip the balance.

In this short article we argued:
1. Computational science is *changing* to be data intensive.
2. Funding agencies should support *balanced systems*, not just CPU farms but also petascale IO and networking.
3. They should allocate resources to support a balanced *Tier-1 through Tier-3 cyberinfrastructure*.